\documentclass[prl,aps,tightenlines,twocolumn,nofootinbib]{revtex4}
\usepackage{graphicx}

\begin{document}

\title{MSSM Higgses as the source of reheating and all matter}

\author{Kari Enqvist$^{a,b}$, Shinta Kasuya$^b$,
        and Anupam Mazumdar$^c$}

\affiliation{
$^a$ Department of Physical Sciences,
     P. O. Box 64, FIN-00014 University of Helsinki, Finland.\\
$^b$ Helsinki Institute of Physics, P. O. Box 64,
     FIN-00014 University of Helsinki, Finland.\\
$^c$ Physics Department, McGill University,
     3600-University Road, Montr\'eal, H3A 2T8, Canada.\\}


\begin{abstract}
We consider the possibility that the dark energy responsible for
inflation is deposited into extra dimensions outside of our observable
universe. Reheating and all matter can then be obtained from the MSSM
flat direction condensate involving the Higgses $H_u$ and $H_d$, which
acquires large amplitude by virtue of quantum fluctuations during
inflation. The reheat temperature is $T_{RH} \lesssim 10^9$~GeV so
that there is no gravitino problem. We find a spectral index
$n_s\approx 1$ with a very weak dependence on the Higgs potential.
\end{abstract}


\maketitle
The large dark energy of the early inflationary universe provides us
with three things: superluminal stretching of space; quantum
fluctuations of scalar fields which may seed density perturbations;
and, once the dark energy decays, the origin of all matter. Although,
conventionally one relates the spectrum of density perturbations to
the properties of the inflaton potential responsible for the early
dark energy, recently it has been realized that this is not a
necessary condition for a successful inflationary scenario. In
curvaton models the dark energy induces quantum fluctuations in a
field whose energy density during inflation is negligible small but
which may later become dominant
\cite{Sloth,Lyth,Others,Enqvist1,Enqvist2,Lyth1}.
When the curvaton decays, its isocurvature perturbations will be
converted to the usual adiabatic perturbations of the decay products,
which thus should ultimately contain also Standard Model (SM) degrees
of freedom. In fact, the logical separation between the field
responsible for density perturbations and the geometrical stretching
of space was already apparent in pre-Big Bang models (for a review, 
see~\cite{Lidsey}).

The curvaton scenario also points towards the possibility that the
inflaton decay products do not necessarily need to give rise to SM particles.
This is an issue that relates to the yet unanswered question of
"how does the inflaton field couple to SM degrees of freedom?".
Indeed,
it has been suggested that the inflaton might not couple to the
observable sector at all but could within the curvaton
framework decay into hidden degrees of freedom~\cite{Enqvist1,Enqvist2}.
In such a case the curvaton field could be found among the flat
directions of the MSSM (for a review, see~\cite{Enqvist02}).
These are described in terms of order parameters, which are
combinations of squarks, sleptons and Higgses, which in the limit
of exact supersymmetry (SUSY) have a vanishing potential. The
MSSM flat directions have all been classified in \cite{gherghetta96}.
Flatness is lifted by SUSY breaking and by non-renormalizable terms
\cite{dine96,gherghetta96}, but during
inflation the MSSM flat directions can be effectively
massless\footnote{This requires the absence of a Hubble-induced mass
term $\sim H$.}  and be subject to quantum fluctuations with a
spectrum identical to the usual inflaton field. The MSSM curvaton, when
it eventually decay, would be a natural explanation for the origin of ordinary
matter.

Note that even if the inflaton field did not decay into SM particles,
it must decay into something.  As a consequence, one would argue that
there must be some inflaton-induced reheating with some resulting
background energy, thermal or non-thermal, that contributes to the
dynamics of the curvaton evolution. Indeed, this is the usual picture
addressed in a number of papers
\cite{Sloth,Lyth,Others,Enqvist1,Enqvist2,Lyth1}. However, with
the advent of the extra dimensions of brane world and string motivated
cosmologies~\cite{Quevedo}, there arises a new possibility:
inflationary dark energy might simply disappear into the bulk.

In the present paper we will consider a scenario in which we live on a
brane (or a stack of branes) where the stretching of space (of our
brane) and the induction of quantum fluctuations of massless fields is
sourced by some dynamics in the bulk.  We further assume that once the
effective dark energy decays, only an insignificant part of the
initial energy density will be deposited on our brane. We will show
that the required adiabatic density perturbations can then be obtained
from the simplest flat directions involving the MSSM Higgses only. It
is obvious that in this case the decay of the curvaton field gives
rise to the quarks and leptons, together with CDM in the form of
lightest supersymmetric particles (LSPs).

We do not have a concrete model for brane induced inflation.  Many
attempts for such a model have been presented in the literature, but
all of them have some problems
\cite{Tye,Burgess,Mazumdar,Rabadan,Shiu,Horace,Kallosh,Brandenberger,Kachru}.
One starting point is the fact that superstring theories admit both
stable BPS Dp-branes~\cite{polchi} and unstable non-BPS branes~\cite{sen1}.
Inflation could then be induced by virtue of attraction between a
brane-anti-brane pair. A pair of brane-anti-brane naturally breaks
supersymmetry which gives rise to a Coulomb-like interaction if the
branes are far apart~\cite{Burgess}. Inflation ends when
brane-anti-brane separation becomes of the order of string scale,
whence the tachyonic instability ends the inflationary epoch. Reheating
due to tachyonic instability primarily reheats the bulk gravitons. The
problems and virtues associated with such a set-up has recently been
discussed in~\cite{Kachru}, but the process of reheating remains
unclear. If the size of the bulk is sufficiently large then the
entire energy of brane-anti-brane annihilation or a decay of an
unstable brane could be absorbed in the bulk in the form of
gravitons~\cite{Maldacena}. They might still appear to us as a
non-thermal background, but if the volume of the bulk is large
enough, there could be a suppression
effect.  If the bulk geometry is warped, the released energy could
possibly be deposited on some other, far-away brane.

Inflation may also occur due to stack of coincident non-BPS Dp branes
\cite{Mazumdar}, which are unstable due to the presence of world
volume tachyons. In this picture the tachyon rolls down and reheats
the bulk degrees of freedom. Inflation can also occur if the two branes
are at an angle, whence the pair of branes could be just Dp. Inflation
again ends when there is a tachyon appearing at a critical angle.
Reheating in this scenario is quite different~\cite{Rabadan,Horace}:
either the branes combine to form a single brane or produce a minimal
energy configuration with a conserved charge. The difference in energy
between initial and final configuration goes into reheating the bulk
degrees of freedom.

A scenario for inflation involving a gas of D-branes embedded in
higher dimensions has also been presented \cite{Brandenberger}.
Inflation takes place because of the heaviest branes in the spectrum
dominate the energy density, which has an equation of state with a
negative pressure. Inflation ends in this picture when correlation
length set by the branes become equal to the Hubble radius. Reheating
in this model occurs primarily due to the decay of a brane into the
bulk. It seems that at least within this class of models, it is
possible to deposit the effective dark energy as seen by the observer
on our brane into the bulk. Another example might be brane world
inflation known as steep inflation \cite{steep}, which has inefficient
reheating, or brane embedded in infinite extra dimensions. In the latter
case the emitted gravitons from the brane can be lost for ever in the
bulk~\cite{dvali}. Yet another interesting scenario could be if the 
inflaton is only charged under some bulk degrees of freedom, and the bulk 
is infinitely large, then the energy density of the inflaton decay 
products can be red-shifted away from the brane~\cite{Lorenzana}.

Although no fully consistent stingy inflationary and reheating model exists,
let us nevertheless assume that after the end of inflation, unlike within
the standard picture of scalar inflation, our brane could remain
essentially devoid of entropy.  While inflation lasts, the effective
dark energy triggers field fluctuations along the (MSSM) flat
directions. In general they cannot be excited simultaneously; rather,
once a condensate forms in one particular direction, the rest are no
longer flat. Hence a typical situation is where one flat direction is
chosen randomly within a single horizon volume which inflates and
becomes the observable universe.

The crucial requirement for an MSSM flat direction to act as a
curvaton is that it does not receive a Hubble-induced mass term during
inflation. This is known to hold true at least in D-term inflation
\cite{binetruy96} and generically in theories with `Heisenberg
symmetry' \cite{gaillard95}. In the latter case a one-loop
contribution eventually gives a Hubble-induced mass correction to the
flat directions (other than stops) of order $10^{-1}H$
\cite{gaillard95}, where $H$ is the Hubble rate during inflation.
There might be other possibilities, too. Let us here just assume that
no Hubble mass term is induced during inflation and that the
energy scale of the inflaton is $V_I \sim H^2M_{p}^2$, where
$M_{p}\sim 2.4\times 10^{18}$~GeV. After inflation we
assume that the energy density goes into exciting the degrees
of freedom residing in the bulk.

The flatness of the potential will be lifted by supersymmetry breaking
and by non-renormalizable terms of the form $W=\lambda\Phi^n/nM^{n-3}$,
where $M$ is a cutoff scale and $n\gtrsim 4$ is the dimensionality of
the non-renormalizable operator; for each flat direction, there exists
a set of allowed non renormalizable operators (see \cite{Enqvist02}).
In general, the flat direction potential can be written as
\begin{equation}
\label{pot}
    V(\phi) = \frac{1}{2}m_{\phi}^2\phi^2
    + \frac{\lambda^2 \phi^{2(n-1)}}{2^{n-1}M^{2(n-3)}}\,,
\end{equation}
where $\Phi=\phi e^{i \theta}/\sqrt{2}$ and the first term comes from
supersymmetry breaking so that $m_{\phi}\sim$ TeV.

Let us now focus on the simplest MSSM flat direction
\begin{equation}
\label{example}
H_u=\frac1{\sqrt{2}}\left(\begin{array}{l}0\\ \phi\end{array}\right)\,,~
H_d=\frac1{\sqrt{2}}\left(\begin{array}{l}\phi\\ 0\end{array}\right)~\,.
\end{equation}
We consider only large amplitudes so that we can ignore the $\mu$-term and the
Higgs mass terms. Then the effective potential for the $H_u H_d$ flat
direction can be written as
\begin{equation}
    V(\phi) = \lambda^2 \frac{|\Phi|^6}{M_p^2}\,,
\end{equation}
where we parameterize the expectation value of the flat
direction as $\Phi(=\phi e^{i\theta}/\sqrt{2})$. Here $\lambda$ is a
constant of $O(1)$ which we set to unity for simplicity.

During inflation, when the cosmological scales leave the horizon, the
curvature of the potential should be small enough for the fluctuation
not to be damped~\cite{Lyth1}. The amplitude of the fluctuation should
also be of the correct order of magnitude. Hence we require that
\begin{equation}
    V''(\phi_*) = \beta^2 H_*^2, \qquad
    \frac{H_*}{\phi_*}=\delta\,,
\end{equation}
where star denotes the value evaluated at the horizon crossing, and
$\beta \ll 1$ and the perturbation $\delta \sim 10^{-5}$. Thus we obtain
\begin{equation}
    \phi_* \sim \beta\delta M_p, \qquad
    H_* \sim \beta \delta^2 M_p\,.
\end{equation}
Therefore we find the scale of inflationary dark energy to be
$V_{I}^{1/4} \sim (H_*M_p)^{1/2} \sim \beta^{1/2}\delta~M_p$.

According to our assumption, after inflation there is no energy
density on our brane except for the condensate along the $H_u H_d$
flat direction. Once inflation is over, the condensate starts to
oscillate and eventually decays. The maximum temperature which can be
achieved after the decay is $T_{max} \sim [V(\phi_*)]^{1/4}$. Notice
that it is much smaller than the scale of inflation. $T_{max}$
corresponds to an instantaneous decay of the flat direction into a
thermal bath of MSSM particles. We will show below that this indeed is
the case.

The Higgs flat direction couples respectively to MSSM fermions and bosons as
$f\phi\psi\psi$ and $f^2\phi^2\chi^2$, where $f$
represents gauge or Yukawa coupling constant. After inflation the
amplitude of the flat direction is very large so one expects both
fermionic and bosonic preheating \cite{preheating}. Preheating in the context
of MSSM flat direction has been discussed in \cite{postmapr}. The
$q$-parameter in this case is given by
\begin{equation}
    q \sim \frac{f^2\phi_*^2}{\omega^2} \sim f^2
    \left(\frac{M_p}{\phi_*}\right)^2 \gg 1\,,
\end{equation}
where $\omega = \sqrt{V''(\phi_*)}$. Typically, the momentum which the
produced particle carries is $k_{res} \sim \omega q^{1/4}$. Since the
occupation numbers of fermion and boson are $n_k^{(f)} \sim 1$ and
$n_k^{(b)} \sim 1/f^2$, respectively, the energy density may be
estimated as
\begin{eqnarray}
    \rho_f & \sim & n_k^{(f)} \omega^4 q \sim f^2 \omega^2 \phi^2
    \sim f^2 \rho_{\phi}\,,
    \nonumber \\
    \rho_b & \sim & n_k^{(b)} \omega^4 q \sim \omega^2 \phi^2
    \sim \rho_{\phi}\,.
\end{eqnarray}
Back reaction terminates the resonant particle production when the
energy density of the produced particle becomes of the order of the
energy density of the oscillating condensate. Notice that there is
less decrease of the energy density due to cosmic expansion during
preheating in this case, because the Hubble parameter is much
smaller than the curvature of the potential for the flat direction:
$H \sim \omega (\phi_*/M_p) \ll \omega$.

Once MSSM fermions and bosons are produced, the energy density of
(non thermal) radiation evolves as
$\rho_{rad}=\rho_f+\rho_b \propto a^{-4}$, while the residual
oscillation of the flat direction decreases as
$\rho_{\phi} \propto a^{-9/2}$, where $a(t)$ is the scale
factor of the universe. After the mass term in the effective potential
of the flat direction dominates, the latter changes its evolution to
$\rho_{\phi} \propto a^{-3}$. The equality of the energy densities
occurs when $\phi=\phi_{eq} \sim (\beta\delta)^{-1} m$, and if the
decay of the flat direction takes place before this time, the
contribution to the total radiation density is subdominant. This happens if
\begin{equation}
    f \gtrsim \sqrt{8\pi}\left(\frac{\phi_{eq}}{M_p}\right)^{1\over2}
    \sim \sqrt{8\pi}(\beta\delta)^{-\frac{1}{2}}
    \left(\frac{m}{M_p}\right)^{1\over2}\,.
\end{equation}
For $\beta=0.1$ and $\delta=10^{-5}$, the right hand side reads
$\sim 10^{-4}$. Since the Higgs has couplings much larger than
$10^{-4}$ (e.g. gauge couplings), the $H_u H_d$ flat direction decays well
before the equality time. Hence the amount of radiation is totally
determined by the preheating era. Notice that the radiation produced
through preheating has non thermal distribution and will take a very
long time to thermalize. Nevertheless, the
evolution of the total radiation energy density is not affected by this.

Let us now estimate the reheat temperature. Since the
maximum temperature is $T_{max} \sim [V(\phi_*)]^{1/4}$,
we may write
\begin{equation}
    T_{RH} \lesssim T_{max} \sim 10^9
    \left(\frac{\beta}{0.1}\right)^{3\over 2}
    \left(\frac{\delta}{10^{-5}}\right)^{3\over 2} {\rm GeV},
\end{equation}
although depending on the thermalization process, the true reheat
temperature could be much lower than $10^9$ GeV. For a generic $n=6$
($n=7$) direction the maximum reheat temperature would be
$10^{13}~(10^{14})$ GeV, so that the avoidance of the gravitino
problem~\cite{Ellis} is not automatic in general. However in the
particular case of the $H_u H_d$ direction there is no gravitino
problem.

Regarding the density perturbations, the amplitude of the fluctuation
created during inflation will be imprinted on
radiation. The spectral index of microwave temperature perturbations
can then be evaluated as \cite{LyWa}
\begin{equation}
\label{nspectr}
    n_s-1 = 2 \frac{\dot{H_*}}{H_*^2}
    + \frac{2}{3}\frac{V''(\phi_*)}{H_*^2}.
\end{equation}
In our set-up, the change of the Hubble parameter is negligible. Hence we find
$n_s-1 \approx 0.007$ for $\beta=0.1$, which gives a very flat
spectrum consistent with recent WMAP observation of $n_s=0.99 \pm 0.04$
\cite{Spergel}.

One should note that the $H_u H_d$ direction has a vanishing baryon
and lepton number. Hence there is no baryonic nor leptonic
isocurvature problem~\cite{McDonald}. There is no Affleck-Dine baryogenesis,
either, but since the reheat temperature $\gg {\cal O}(1)$ TeV,
baryons could be produced by sphaleron processes at the electroweak
phase transition.

In conclusion, we have pointed out that in theories with extra
dimensions there may exist the possibility that the effective dark energy
which drives inflation could be deposited outside of our observable
brane. As a consequence, reheating and all matter could originate
simply from a MSSM flat direction involving the Higgses, which gets
excited during inflation. Moreover, as we showed, the reheat
temperature is low enough to avoid the gravitino problem.

It is interesting to note that the spectral index Eq. (\ref{nspectr})
depends -- albeit very weakly -- on the Higgs potential and hence on
the properties of Higgs particles that could, at least in principle,
be determined in the laboratory. Such connections between the
properties of elementary particles and temperature fluctuations of the
microwave sky would naturally be highly desirable.  It remains to be
seen whether inflationary models based on string theory can allow for
the depositing of the dark energy outside of our brane as required in         
the present framework.

\vskip5pt
K.E. is supported partly by the Academy of Finland grant no. 75065, and
A.M. is a CITA-National fellow and acknowledges Cliff Burgess,
Jim Cline, Guy Moore, Horace Stoica, Mahbub Majumdar and Abdel Lorenzana
for helpful discussion.

\vskip20pt


\end{document}